# Using Computational Approaches in Visual Identity Design: *A Visual Identity for the Design and Multimedia Courses of Faculty of Sciences and Technology of University of Coimbra*

Sérgio M. Rebelo[1] (srebelo@dei.uc.pt); Tiago Martins[1] (tiagofm@dei.uc.pt)
Artur Rebelo[1]; João Bicker[1]; Penousal Machado[1]

[1] University of Coimbra, DEI, CISUC

**Abstract**

Computational approaches are beginning to be used to design dynamic visual identities fuelled by data and generative processes. In this work, we explore these computational approaches in order to generate a visual identity that creates bespoke letterings and images. We achieve this developing a generative design system that automatically assembles black and white visual modules. This system generates designs performing two main methods: (i) *Assisted generation*; and (ii) *Automatic generation*. *Assisted generation* method produces outputs wherein the placement of modules is determined by a configuration file previous defined. On the other hand, *the Automatic generation* method produces outputs wherein the modules are assembled to depict an input image. This system speeds up the process of design and deployment of one visual identity design as well as it generates outputs visual coherent among them. In this paper, we compressively describe this system and its achievements.

**Keyword**

Computer-aided design software; Dynamic visual identity; generative design; graphic design;

# 1. Introduction

Novel digital technologies have significantly changed the contexts and the tools of graphic designers' practice. However, most of the graphic designers do not have change yet their practice (Armstrong & Stojmirovic, 2011; Blauvelt, 2011; Dubberly, 2008). They continue to use computational approaches as tools rather than explore the potential of them in order to develop novel ways of to communicate with the audience (Bohnacker, Groß, Laub, & Lazzeroni, 2009; Maeda, 2004; Reas, McWilliams, & LUST, 2010; Richardson, 2016).

The use of these computational approaches allows graphic designers to explore novel visual and conceptual possibilities enabling the development of systems that craft processes that generate multiple outcomes, instead of crafting every single outcome (Bohnacker et al., 2009; Reas et al., 2010). Also, the resulting outcomes have mutable and flexible characteristics that enable the design of more customisable artefacts. This way, they might create closer relationships with their audience (Armstrong & Stojmirovic, 2011; Blauvelt, 2008; Bohnacker et al., 2009; Richardson, 2016). Although these characterise are not restricted to the digital artefacts and also appears in print artefacts (e.g. (Neves, 2016), (Moniker, 2013) or (Eatock, 1998)), computational tools appear to be the primary tools and the driving force to the emergence of these kinds of artefacts (Armstrong & Stojmirovic, 2011; Blauvelt, 2008).

The recent emergence of the dynamic visual identities (or dynamic corporate brandings) are clear examples of this change in basic assumptions of Graphic Design. During the last years, the classic principles of the visual identity were beginning to be called into question. Normally, the process of design one visual identity is about translating the institution's cores values into a visual symbol that describes the desired interaction between the institution and the general audience (Erlhoff & Marshall, 2008). Nevertheless, the traditional, static and uniforms, logotypes no longer could represent the complexity of the institutions of nowadays (Blauvelt, 2011; Felsing, 2010; Martins, Cunha, Bicker, & Machado, 2019; van Nes, 2013). This way, graphic designers began exploring computational approaches to create dynamic visual identities systems capable of handling with, for instance, real-time data input and generative processes (Armstrong & Stojmirovic, 2011).

In this paper, we explore how computational approaches could be employed to generate a dynamic visual identity. This work is aligned with the design of a new visual identity for the bachelor and master's degrees in Design and Multimedia of the Faculty of Sciences and Technology of University of Coimbra. The main aim was to develop a visual identity that translates the interdisciplinary between information technologies, interaction and graphic design existing in the study programme of this course. This way, we decide to present a visual identity as a computational system to fulfil three main requirements. First, the visual identity system should create dynamic and flexible outputs. Second, the system should enable visual identity needs to be deployed to multiple media and formats in a faster and easier way. Finally, the system should promote easier interactions and closer communication with the visual identity artefacts and their audience.

The resulting visual identity system explores the modularity as its main design characteristic, enabling the development of distinctive and, at the same time, consistent designs (Lupton & Phillips, 2015). This way, it is characterised by the use of a specific set of black and white visual modules assembled in order to create lettering and images. We develop a computational system that employs generative design techniques to automatically perform this task. The system employs two main methods to generate designs: (i) *Assisted generation*; and (ii) *Automatic generation*. *Assisted generation* allows generating visual compositions based on the data from a previously defined configuration file. This method is specially designed to create variations of logotype in an effortless and dynamic way. On the other hand, the *Automatic generation* method produces visual compositions wherein the assemble of visual modules is based on the brightness of an input image. This way, the resulting composition, in their whole arrangement, depicts an input image. Thus, the system could generate multiple kinds of visual images that, although distinct, maintain the visual aesthetics of visual identity, in an automatic manner.

The remainder of this paper is organized as follows. Section 2 summarises some related work. Section 3 explains how our system works and presents some preliminary results. Finally, section 4 presents our conclusions and points the direction for future work.

**2. Background**

A visual identity is much more than a trademark. It is defined by a set of structural visual elements that, at least, include an identifying symbol (e.g. logotype, mark, or trademark), a set of colours and a typeface (van Nes, 2013). Symbols had always been utilised for visual identification of institutions. Nevertheless, the increase of the scope of many institutions and companies, during the 1950s, boosting the emergence of visual identity systems in the sense that we know nowadays. The work of graphic designers and design firms such as Paul Rand, Saul Bass, Otl Aicher, Muriel Cooper and Chermayeff & Geismar are cases in point (Meggs & Purvis, 2012). However, these visual identity systems were designed to unify the communication of institutions and, thus, they were solid and unwavering marks (Armstrong & Stojmirovic, 2011; Blauvelt, 2011; Meggs & Purvis, 2012).

Nowadays there still is a demand for the creation of visual identities. However, traditional static and uniforms visual identities no longer can transmit the core values of contemporary institutions. This way, graphic designers have begun to explore the creation of dynamic visual identities characterised by variability, context-relatedness, processuality, performativity, and non-linearity (Armstrong & Stojmirovic, 2011; Blauvelt, 2011; Felsing, 2010; Martins et al., 2019; van Nes, 2013). Moreover, the recent technological advancements, along with the democratisation of the Internet, allowed graphic designers to amply explore the possibilities of dynamic visual identities in a manner that are not practicable at years ago (Armstrong & Stojmirovic, 2011; van Nes, 2013).

As we know, the first dynamic graphic mark is used in books of Alfred A. Knopf publishing house, since 1915. Each book edited by this publisher is printed with a different variation of a borzoi graphic mark. During the 1950s, Karl Gerstner presented how the concept of flexibility could be employed in visual identities in his design to *Boîte a Musique.* He presented this identity as a system that could be adapted to different functions and proportions maintaining the visual style and personality (Gerstner, 1964). In the 1980s, Manhattan Design studio created the MTV logotype, displaying that a logotype could take multiple shapes and variations, maintaining always its personality (Meggs & Purvis, 2012).

Nevertheless, it is with the turn of the millennium that there is an increase of the number of dynamic visual identities (Blauvelt, 2011; Felsing, 2010; Martins et al., 2019; van Nes, 2013). There are well-known, for instance, the identity for the Scandinavian peninsula of *Nordkyn,* designed by Neue Studio (Neue, 2010), the *Casa da Música* identity and its auxiliary colour picker application by Sagmeister, Inc. (Sagmeister & Walsh, 2016), the *AOL identity,* by Wolff Olins (Wolff Olins, 2007), among others. (Martins et al., 2019) presents a good overview of the field. However, we are only interested, in the context of this paper, in study dynamic visual identities that explore the combination of graphics in a modular way. Following, we will present some works.

*Walker Expanded* (Blauvelt, 2011; Felsing, 2010) designed by Andrew Blauvelt and Chad Kloepfer, in 2005, is one of the firsts well-successful dynamic visual identity system. It is a flexible system, based on vertical stripes, that enables the use of different words, partners, motifs, and colours always keeping the efficiency and consistency across institution communications. The system operates as a system font. However, instead of individual characters, it contains words and customisable patterns that can be merged in the same line.

*Monadnock* identity (Catalogtree, 2008) designed, in 2008, by Catalogtree studio with collaboration with the computer programmer Lutz Issler, is a visual identity that reinvents themselves at each interaction. Each time that the logotype is saved, exported, or printed, a novel version of itself is randomly generated using two or more of the characters of the company's name.

IDTV identity system designed by Lava, in 2007, (Lava, 2007; van Nes, 2013) is defined by the multiple combinations of four visual modules (or pixels). Each one is an IDTV business unit (tv, film, documentary, and events). The logotype is created through the combination of these modules, in different scales and colours, in a predefined grid. Furthermore, the modules are used to create visual applications of identity in multiple media and formats. This way, whatever it is designed with these modules, it is always visually related to IDTV.

In 2009, Mind Design studio, in collaboration with Simon Egli, designed a dynamic visual identity system for the model agency Tess Management (Mind Design, 2009; van Nes, 2013). The identity is composed of several logotype variations generated through the placement of visual modules, inspired in art-deco elements, over a grid of squares and half-squares. Moreover, the same elements are used to create decorative frames.

Visual identity for AGI Congress 2010 held in Oporto, designed by R2 studio, also explores the modularity in the context of dynamic visual identities (R2 Design, 2010). One of its main characteristics is the use of modular compositions generated using visual modules inspired in typical Portuguese tiles — the *Azulejos*. These compositions were generated using a bespoke system, developed in collaboration with Penousal Machado, that assembles modules in different scales and positions in order to create an output that depicts an input image. Moreover, the system creates abstract compositions of modules (i.e. patterns) that when applied to any media related them for AGI Oporto 2010.

In 2012, Lava studio designed a dynamic visual identity to the Moscow Design Museum (LAVA, 2013). The identity was developed as of one hard-angled grid based on the traditional Russian crystal pattern. From this grid, they designed dozens of visual modules and designed a dynamic visual identity. The final result was a highly recognisable and flexible visual identity that can be transposed to multiple media, sizes and formats.

The *Lille Métropole* dynamic visual identity proposal designed by a team of designers from Maryland Institute College of Arts, in 2013, used visual modularity as the main design characteristic (Lupton, 2014; Tabet, Blake, Sherwood-Forbes, & Zhou, 2013). They designed and defined a visual element for each city within the metropolitan area. Beyond, they create four visual modules to represent each city through operations of scale and rotation on the designed elements. The visual identity is the combination of these modules over a rectangular grid. These combinations work in two ways: (i) they could identify a city when only modules related to the city element are placed; or (ii) they could identify the whole metropolis when modules designed as of the different elements are combined. The proposed system proved to be extremely flexible when applied to various media, from print media to digital media and environmental signage.

In 2014, a visual identity for the celebrations of the thirtieth anniversary of the Informatic Engineering degree of the University of Coimbra and the twentieth anniversary of its department was developed by (Matos, Bicker, & Cruz, 2014). In this visual identity, the several ways of representing a pixel, along the years, are used as the main concept to design a set of shapes (i.e. squares, circles and rectangles). These visual shapes are, then, combined, over the same grid, with different colours and sizes to generate letterings and patterns.

## 3. Approach

Visual identities are defined by a set of structural elements that, at least, include an identifying symbol, a set of colours and a typeface. Dynamic identities work in the same way. In this paper, we describe the approach to develop the computational system that automatically assembles the set of visual modules according to the visual identity aesthetics and standards. The set of visual modules include nine black and white squares (see Figure 1). The modules were previously designed and loaded in the system as vector files (SVG format). The system employs a set of generative process to combine the modules. Outputs are rendered through two distinct methods: (i) *Assisted generation* (see subsection 3.1); and (ii) *Automatic generation* (see subsection 3.2). The user at the start of the generation process should choose one. The next subsections comprehensively describe how each method is performed by the system.

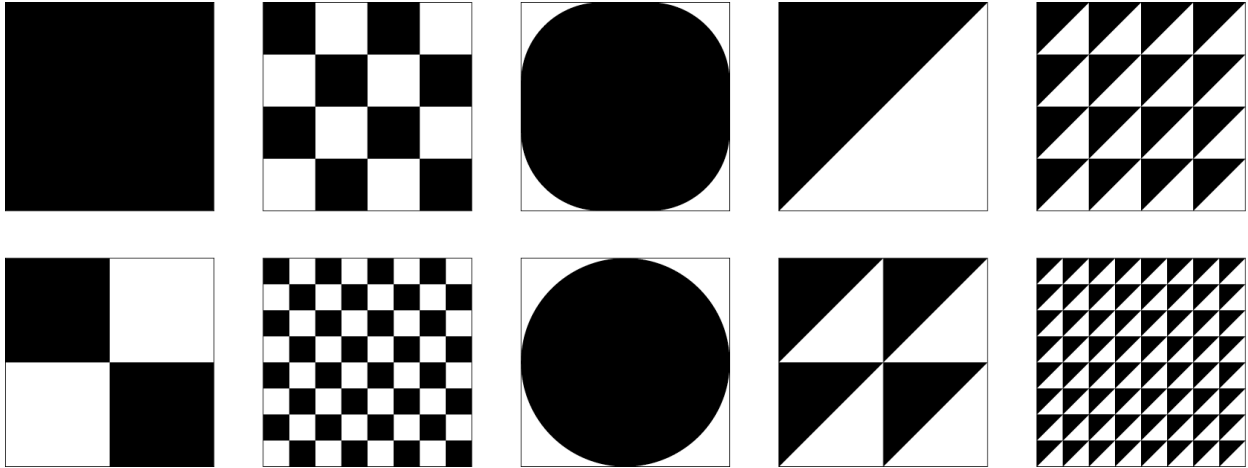

Figure 1: Set of visual modules basis from the visual identity..

However, the visual identity goes beyond the generative system. The text that goes along with the identity is always composed using the combination of two typefaces: Helvetica Neue LT Regular, designed by D. Stempel AG team as of the Helvetica original designs of Max Miedinger (1983); and Neue Pixel Grotesk designed by M35 (M35, 2018). This identity is designed to mostly work at black and white, albeit we can use any complementary colour or replacing the black or the white by another colour. Figure 2 displays some examples of merchandising designed using this visual identity.

3.1. Assisted Generation

Normally, a dynamic identity is based on a dynamic logotype. This identity is not an exception. We considered important that each course employee and service have their own version of logotype. This way, we endow the system with a method that easier generates versions of the logotype whenever they are necessary. The method generates logotypes performing an assisted placement of modules in canvas based on a configuration file previous defined.

The generation process begins with the definition of the canvas size by the user. Subsequently, the system decides what will be the logotype format. If the user selected a width greater than the height (i.e. landscape format) the system will generate a horizontal version of logotype (see Figure 3Figure 3). On the other hand, if the selected height is greater than the width (i.e. portrait format) the system will generate a vertical version of the logotype (see Figure 4). If in any chance the selected width is the same that height (i.e. square format) the system will generate a vertical version of the logotype.

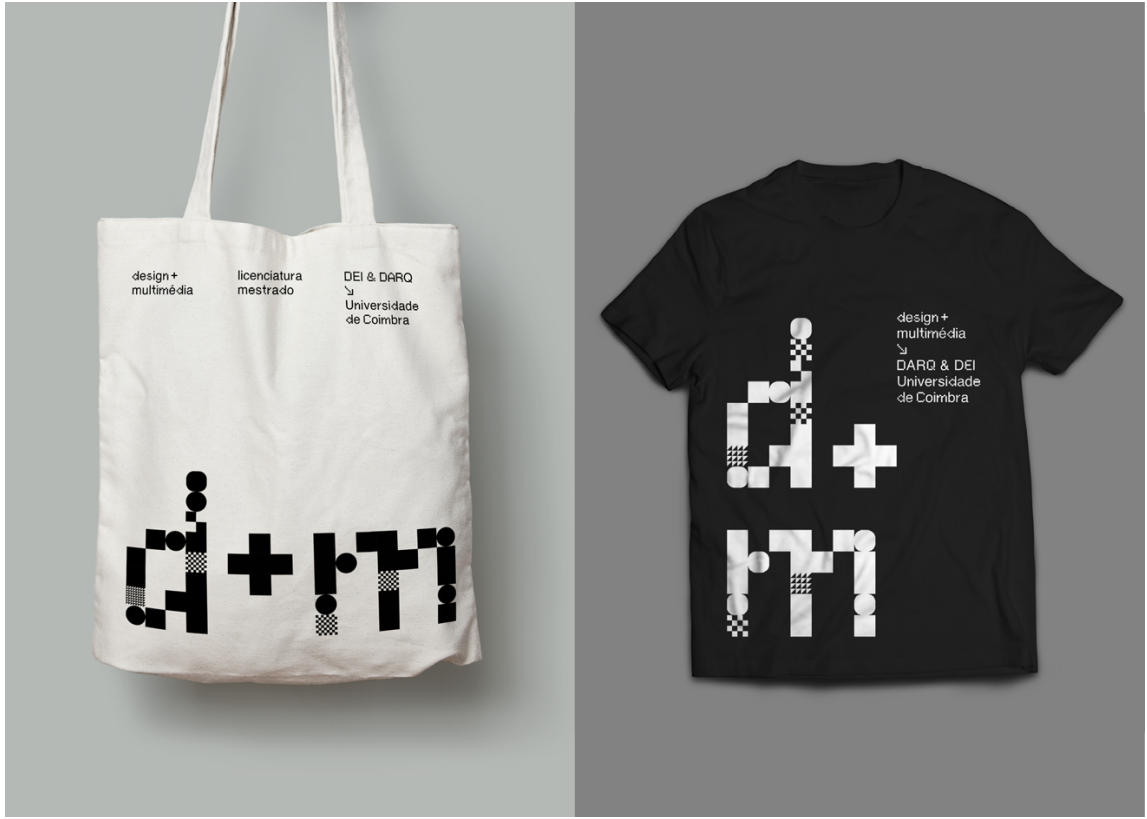

Figure 2: Mock-ups of the visual identity applied in order to design merchandising bags (at left) and t-shirts (at right).

The system places the visual modules according to the information from a configuration file text file. In this text file, the placement settings for each logotype version are described. This way, the system loads a different configuration file whenever it needs to generate a different logotype version. Each file is fulfilled with multiple sequences of characters. Each line in the file represents a row in canvas and the position of each character in text line a column. This way, each character in the text file is a cell in the canvas grid. Each character may have eleven different values. When its value is zero (0) the system ignores the correspondent cell in canvas, i.e. do not place a module in this cell. On the other hand, when its value is an asterisk (*) the system randomly chooses a module to place it the respective cell. Furthermore, the characters could point to a specific module. When the character value is a hexadecimal number between 1 and A, it places a module according to the order displayed in Figure 1Figure 1. In this figure, module '1' is the upper-left module and module 'A' the bottom-right module.

The method begins by subdividing the canvas in a modular grid. The number of columns corresponds to the length of the first line in the text file. The number of rows corresponds to the number of text lines in the document. Subsequently, it runs each cell in the grid and, based on data retrieved by the configuration file, it places the respective module in canvas. In the end, it exports the generated output to raster (PNG file) and vector (SVG file) formats. Figures Figure 3 to Figure 6 display some outputs generated by this method.

Also, we implement in this method some techniques in order to prevent failures in the drawing process. First, it always considers the number of columns based on the first line of the text file. This way, if a line length of a text line is bigger than the maximum number of the columns, it ignores the characters that overpass this value. Besides that, if the line is shorter than the number of columns, it considers that the missing characters are '0' (i.e. blank spaces). Second, it considers that every line in the text file is a row. Therefore, if a line is blank it considers that all positions are '0' in this row (i.e. a blank row in canvas). Finally, if it reads a character that does not have a module defined (bigger than 'A'), it considers that this character is an asterisk.

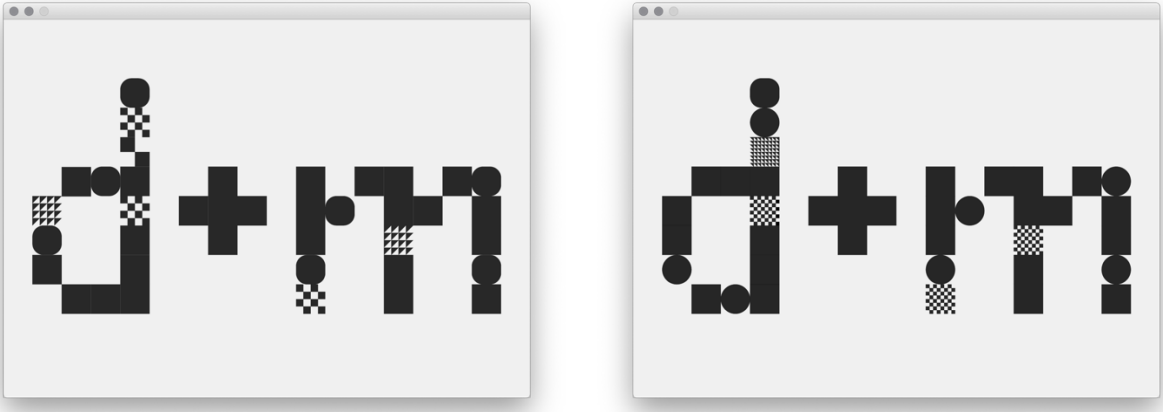

Figure 3: Two horizontal versions of the logotype generated using the same canvas size and configuration file.

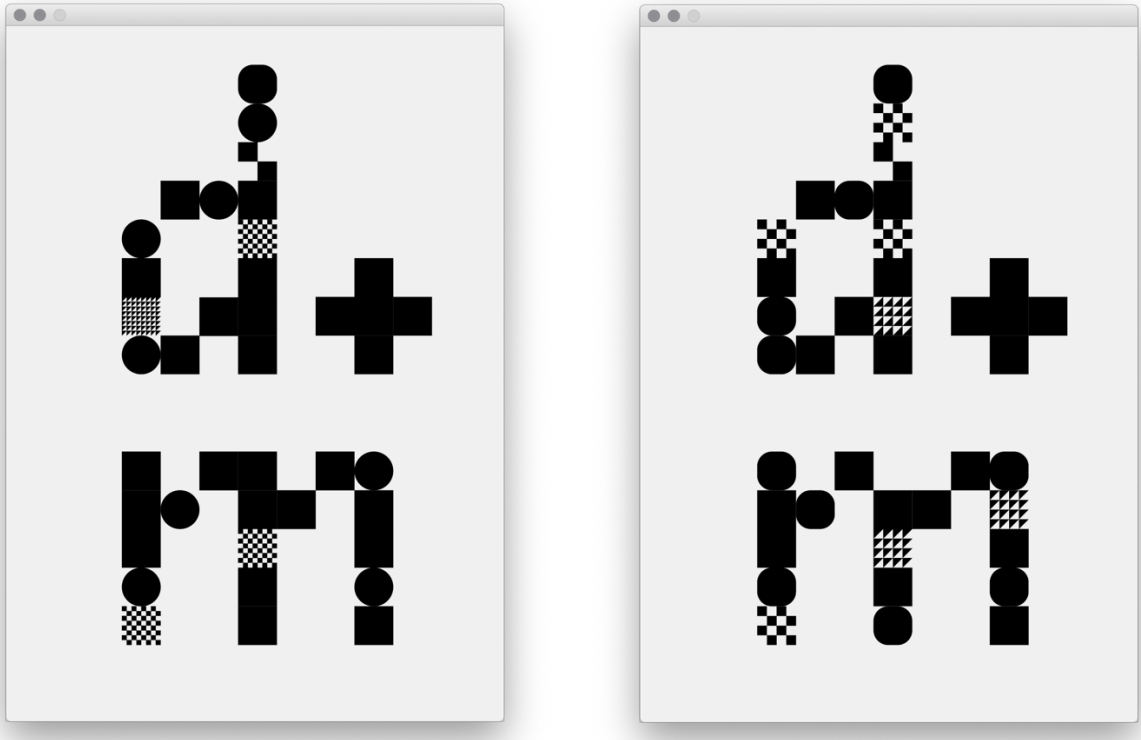

Figure 4: Two vertical versions of the logotype generated using the same canvas size and configuration file.

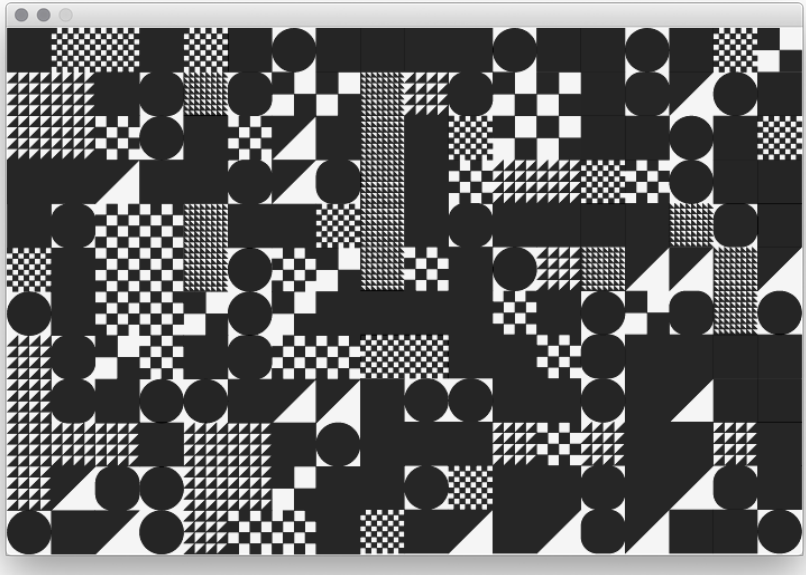

Figure 5: One pattern generated using a randomly generated configuration file.

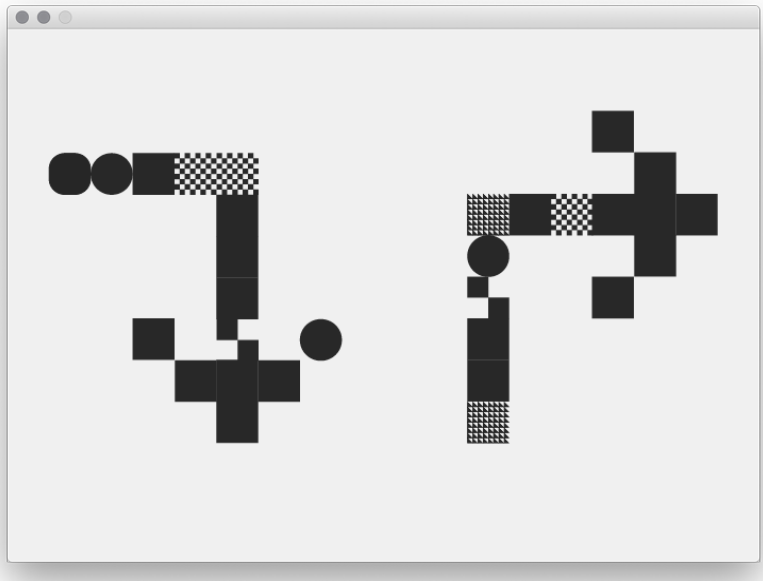

Figure 6: Arrow designs generated using a configuration file created by the user.

This process was created with the main aim to generate logotypes. However, due to this flexibility, it enables the user to generate other kinds of visual compositions such as patterns, letterings and visual symbols and icons. The user might perform this by loading different configuration files. This is a workable solution to easily generate basic icons or patterns in the visual aesthetics of visual identity. However, it could be difficult to design a configuration file to generate more complex compositions. At the end of each generation, the system allows the user to save the configuration file used in generation. Also, it allows the exportation of a blank configuration file template for the user to easily create configurations files. Figures Figure 5 andFigure 6 display some outputs generated using modified configuration files.

3.2. Automatic generation
The system also has the ability to generate visual compositions where the visual modules are assembled in order of depicting an input image given by the user. This way, it automatically generates multiple variations, according to the visual identity aesthetics and standards, based on an input image. Thus, users can generate several design variations until they are satisfied and apply them to different media. On the other hand, this method also allows the generation of visual identity applications that have the potential to create closer relationships with their target people. For instance, it enables the automatic generation of customised designs that could be faster distributed over web pages. *Automatic generation* method achieves this looking to brightness of input image and comparing these values with the brightness of each module. This way, the modules are placed in order to its whole arrangement depict the input image.

To run this method, the user should input an image and define the number of the row that the image will be subdivided. It based on the inputted image's width and in inputted number of rows define the number of columns, once the modules are squares. The process that creates the composition comprises the following steps: (i) convert the input image to greyscale; (ii) normalise each pixel brightness according to a minimum and maximum threshold; (iii) perform subdivision of the image according the modular grid defined previously; (iv) calculate for each grid module the average brightness the pixels located in each ninth part of it; and (v) place the modules inside it according to the computed brightness. Figure 6 displays the image processing process before the placement of the modules. Figure 7 displays the image processing stages before the placement of the modules.

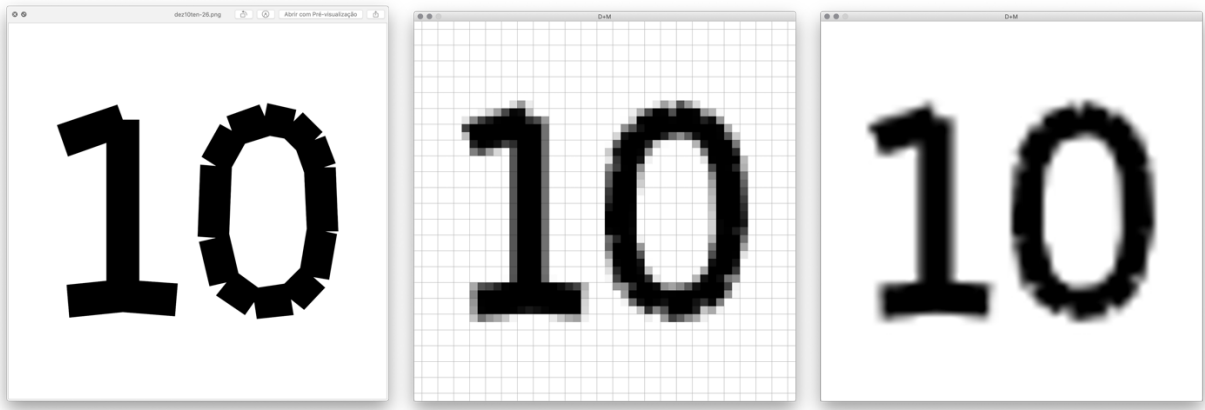

Figure 7: Image processing stages of the input image. From left to right: original image; image convert to greyscale; image normalised according to a preset minimum and maximum thresholds. Input image generated using the system developed by (Pereira, Martins, Rebelo, & Bicker, 2019).

The position of the modules over the grid is implemented in order to it reflects the brightness of the original image. This way, a darker area in the image has more probability of being replaced by a dark module than a bright one. It is achieved by looking not only to the average brightness of the pixels located in a grid section but also to this distribution. In this case, the system subdivides each module into four subdivisions and, after, calculate the average brightness for each subdivision. Performing this task, the system places the modules by the local average of the brightness of the pixels, creating, therefore, smoother results. The average brightness of each pixel was calculated previously by performing the same task to the base visual modules and loaded these values in the system.

This way for each cell in the grid above a preset minimum threshold, we calculate the following aspects mentioned above and initiate the placement of modules sequentially. For each cell in the grid, the system calculates the difference between the average brightness of each one subdivision and the correspondent subdivision in all the modules loaded in the system. In the end, it chooses the most similar module. The most similar module is the one who the sum of all differences is smaller.

If the system places the modules directly in canvas, the outputs generated, from the same image and grid configuration, are always the same. Nevertheless, we are interested to create outputs that although similar are not identical. This way, we introduced some uncertainty in the moment of placement of the modules. This way, the system decides what will the module placed employing a fitness proportionate selection method, i.e. a roulette wheel selection. This method is performed as followed. For each grid cell, the system assigns a probability to each module based on the similarity between brightness. Thus, the modules most similar to the target pixels, in term of brightness, have higher is the probability of being selected. Next, it randomly chooses one module and placed it in composition. In the end, it presents the generated composition to the user and exports the outputs to file in raster (PNG file) and vector (SVG file) format. Figure 8 displays some example outputs. More outputs could be seen in the video available at [www.cdv.dei.uc.pt/2019/dm-dvi/results.mp4](www.cdv.dei.uc.pt/2019/dm-dvi/results.mp4).

Although this method was developed to work with images, we also implemented a video mode. In this mode, it continually generates compositions based on the frames from a video (or video capture) and display this for the user. An example capture video is available at [www.cdv.dei.uc.pt/2019/dm-dvi/capture.mp4](www.cdv.dei.uc.pt/2019/dm-dvi/capture.mp4).

As mentioned above, the *Automatic generation* method enables easier deployment of visual identity applications in various media with multiple sizes and formats. This method presents some advances, compared with the *Assisted generation* method (see subsection 3.1), especially in the design of more visually complex designs. For instance, the designer does not need to define, previously, the arrangement of the modules because the system will automatically choose them, based in the colours of the input image. Also, the designer could use his/her favourite tool to generate an input image in the format and style that he/she desires and, after, uses this method to generate one, or many, outputs according to visual aesthetics and standards. Such as the *Assisted generation* method (see subsection 3.1), since the outputs are exported in vector and raster file, the designer may use his/her favourite tool to make the necessary adjustments in the design.

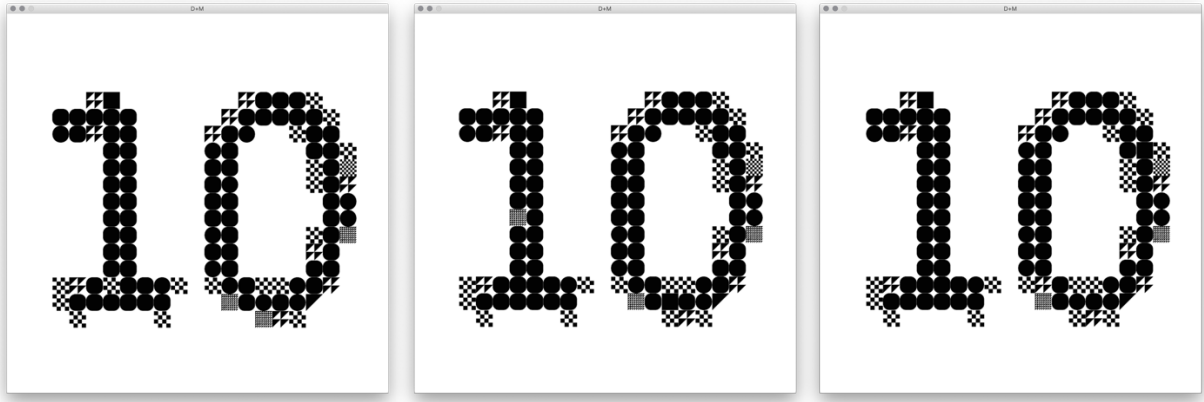

Figure 8: Three outputs generated using the same grid configuration and the same input image. More results could be seen in the video available at www.cdv.dei.uc.pt/2019/dm-dvi/results.mov. Input image generated using the system developed by (Pereira et al., 2019).

On the other hand, this method has the potential for the development of interesting brand activation strategies that promotes the creation of stronger relationships between the visual identity and its target audience. This way, everyone can be related to this new visual identity, stylizing, in the aesthetic of visual identity, the images that like. For instance, we could create interactive installations constantly generates the environment near it (i.e. a digital mirror) and generating the snapshots of student and faculty for identification cards and website.

## 4. Conclusions and Future Work

We have presented a dynamic visual identity for the bachelor and master's degrees in Design and Multimedia of the Faculty of Sciences and Technology of University of Coimbra. This visual identity explores the modularity as its main design characteristic using a set of black and white modules assembled in order to create letterings and images.

The assembling of the modules is made using a bespoke computational design system. The system implements two main generative methods to achieve this: (i*) Assisted generation*; and (ii*) Automatic generation*. The first method designs the modules in a canvas using data retrieved from a configuration file. It is specially designed to design new logotype versions of the logotype, although it could be used for generating other kinds of outputs (such as icon and patterns generation). On the other hand, the

*Automatic generation* method designs compositions when the visual modules in their whole arrangement depict an input image. It performs this by positioning the patterns according to the brightness of the input image. This method is specially designed to create visual identity visual applications. However, it allows an easy generation of all kind of images (or videos) in visual identity aesthetics and standards.

We can identify some advantages of the developed system: (i) it permits easily generation of new versions of the logotype; (ii) it simplifies the process of design dynamic visual identities' applications (the designer only needs to input an image with the desired shape and it generates a similar image in visual identity aesthetic); and (ii) it creates tools to develop a more closer relationship between the institution and its audience and employees.

We can identify some advantages of the developed system: (i) it permits easier generation of new versions of the logotype; (ii) it simplifies the process of design dynamic visual identities' applications; and (ii) it enables the development of artefacts that promote a more closer relationship between the course coordination and its faculty and students.

Future work will focus on: (i) design/generate more visual modules in order to create sharper compositions; and (ii) study the interaction of the target people with visual identity materials.

## 5. Acknowledgements

This work is partially supported by national funds through the Foundation for Science and Technology (FCT), Portugal, within the scope of the project UID/CEC/00326/2019. The first author is funded by FCT under the grant SFRH/BD/132728/2017.